\documentclass[11pt]{article}  
\usepackage{epsfig,graphicx,graphics,amsmath}  
\input epsf  
  
\newcommand{\beq}{\begin{equation}}  
\newcommand{\eeq}{\end{equation}}  
\newcommand{\ben}{\begin{enumerate}}
\newcommand{\een}{\end{enumerate}}
\newcommand{\bitem}{\begin{itemize}}
\newcommand{\eitem}{\end{itemize}}
\newcommand{\bfig}{\begin{figure}}
\newcommand{\efig}{\end{figure}}
\newcommand{\bcen}{\begin{center}}
\newcommand{\ecen}{\end{center}}

\newcommand{\delete}[1]{}

\title{Galactic Rotation Described with Various Thin-Disk Gravitational Models}
\author{James Q. Feng and C. F. Gallo \\ \\ Superconix Inc, 2440 Lisbon Ave, Lake Elmo MN 55042}

\begin{document}
\pagestyle{myheadings}

\maketitle

\bcen
\Large{\bf Abstract}
\ecen

Mature spiral galaxies consist of stars, gases and plasma distributed approximately in a thin disk of circular shape. The rotation velocities quickly increase
from the galactic center and then achieve a constant velocity from the core to the periphery.  
The basic dynamic behavior of mature spiral galaxies are well described by simple models balancing Newtonian gravitational forces against the centrifugal forces associated with a rotating thin axisymmetric disk. 

The Freeman model assumes a mass density distribution 
decreasing exponentially with radius in a thin disk extending to infinity.  
This assumption of exponentially decreasing mass was inspired by the observed exponentially decreasing luminosity curves. 
This exponential mass distribution produces rotational velocity profiles which increase from the galactic center to a maximum toward the outer core (as observed), but then decreases out to the periphery. This assumed exponential distribution is unable to adequately describe the 
measured constant (flat) velocities from the outer core to the periphery. 
By contrast, Mestel's finite disk model assumes a mass distribution decreasing more slowly (inversely with radius) that results in a constant rotational velocity across the entire disk.
But this inverse radius dependence of the mass distribution has an unrealistic singularity at the galactic center and does not describe the central core rotation properly. 
Thus we propose combining the Freeman and Mestel disk models to utilize their strengths and eliminate their deficiencies.   
We utilize the Freeman formula for the central core region, and then the Mestel formula beyond the central core to the galactic rim to describe the mass density distribution in a thin axisymmetric disk.
Employing a connecting scheme for the transition from the Freeman to Mestel formulas, 
this combined model produces rotation curves comparable to the measurements for mature spiral galaxies.   

For a more general thin-disk model, we develop an alternative computational method to solve for the mass density distributions for various given measured rotation curves, instead of calculating the rotation curves from assumed mass density formulas (as assumed by Freeman and Mestel).
We employ a thin axisymmetric disk of finite radius, and then compute radial mass densities that  balance the Newtonian gravitational forces against the centrifugal forces (due to galactic rotation) at each and every point. 
The computational solutions for typical rotation curves show 
corresponding mass density distributions which decrease approximately exponentially in the central core (similar to the Freeman formula), 
and then transition to a slower inverse radius decrease (similar to the Mestel formula) to the periphery.   
   
Thus these diverse approaches yield similar results and are mutaully self-consistent with Newtonian gravity/dynamics.     
   

\section{Introduction} 

\subsection{Observational Knowledge of Radial Galactic Rotation Profiles}
Telescopic images of mature spiral galaxies reveal most of the stars, gas and plasma reside in an approximately circular disk that is very thin compared with its radius (Refs.\cite{BT}-\cite{Bennett}).
The data on galactic rotational velocity profiles 
(Refs.\cite{Rubin1}-\cite{deBlockMH}) of mature spiral galaxies are 
characterized by a rapid increase from the galactic center, reaching a nearly constant
velocity from the outer core to the outer periphery.  
This basic feature may be idealized as  
\beq \label{eq:smooth-V}
V(r) = 1 - e^{-r / R_c} \, , 
\eeq
where $V(r)$ denotes the dimensionless rotational velocity
measured in units of a characteristic velocity $V_0$ and 
$r$ the radial coordinate from the galactic center.
The parameter $R_c$ is a description from the data of the various ``core'' radii of different galaxies. Typical galactic rotational profiles described by (\ref{eq:smooth-V})
are displayed in Fig 1. As indicated by the measurement data, 
the rotation velocity typically rises linearly from the galactic center 
(as if the local mass was in rigid body rotation), and then reach an approximately constant (flat) velocity out to the galactic periphery. 

The observed galactic rotation curves (Eq(\ref{eq:smooth-V}) and Fig.1) 
can not be explained by simply applying the so-called 
{\em orbital velocity law}, derived for a spherically symmetric gravitational field
which is applicable to the Keplerian rotation of our solar-planet system, but not disk-like galaxies.
In fact, the galactic mass distribution calculated by 
the {\em orbital velocity law} applied to these constant (flat) galactic rotation curves 
yields an {\em increasing mass with radius},  
contrary to the measured galactic luminosity curves which decrease exponentially with radius. 

\begin{figure}[htb]
\resizebox{!}{1.25\textwidth}
{\includegraphics{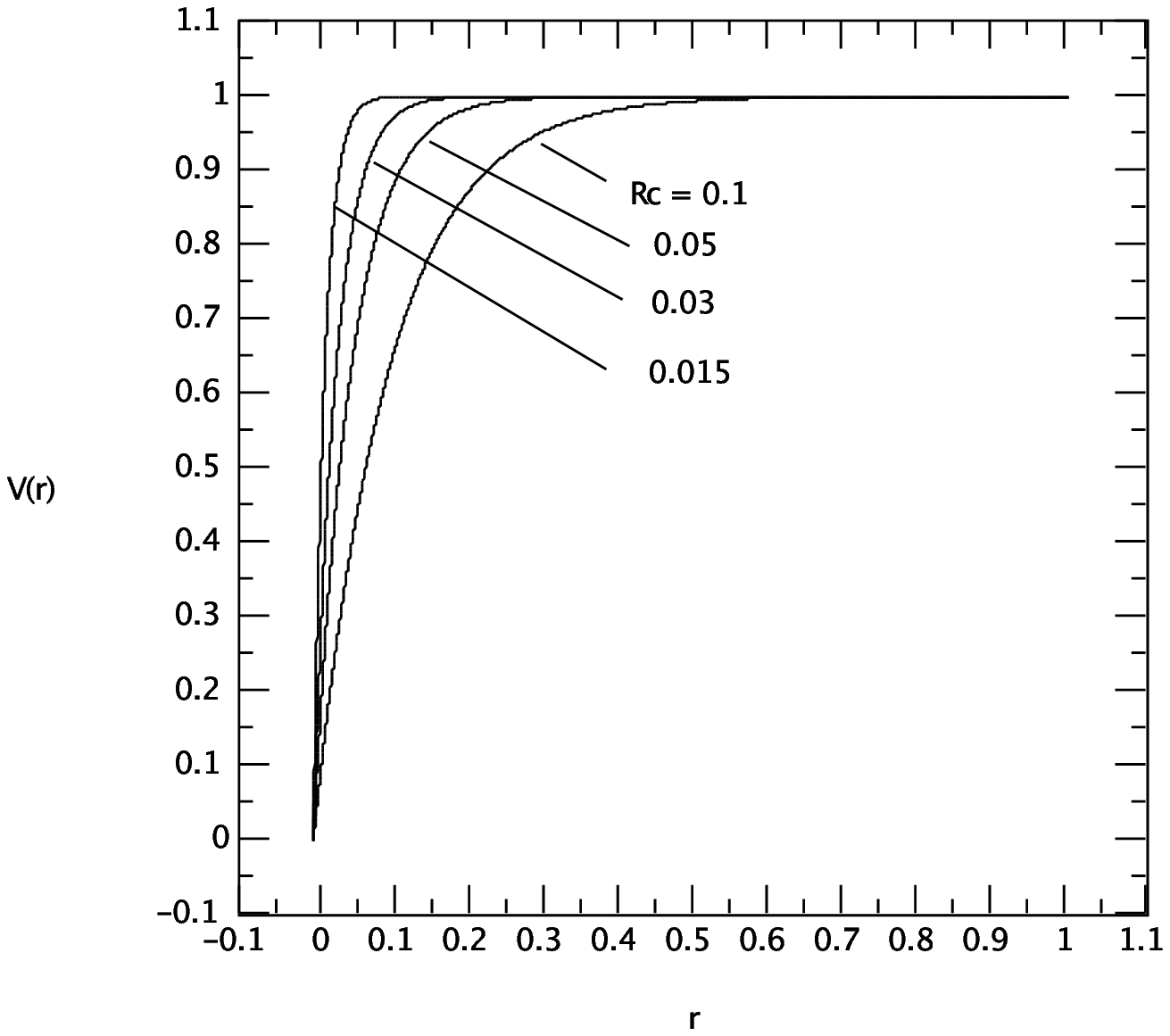}}
\caption{Galactic rotational velocity profiles $V(r)$
idealized from measurements according to 
(\ref{eq:smooth-V})
for $R_c = 0.015$, $0.03$, $0.05$ and $0.1$.}
\label{fig:fig1}
\end{figure}

\subsection{Thin-Disk Gravitational Models} 
For a thin rotating galactic disk (revealed by telescopic observations),  
a balance can be achieved between the 
Newtonian gravitational forces and centrifugal forces at each and every point.  
Because the gravitational field of a thin disk is not spherically symmetric, 
the {\em orbital velocity law} is not applicable.
As presented by Feng \& Gallo \cite{FengGallo1}, 
the equation of force balance in an axisymmetric thin disk can be written as
\beq \label{eq:force-balance0}
\int_0^1 \left[\int_0^{2 \pi} 
\frac{(\hat{r} \cos \phi - r) d\phi}
{(\hat{r}^2 + r^2 - 2 \hat{r} r \cos \phi)^{3/2}}\right] 
\rho(\hat{r}) h \hat{r} d\hat{r}
+ A \frac{V(r)^2}{r}
 = 0 \, ,
\eeq
where all the variables are made dimensionless by measuring lengths (ex., $r$, $\hat{r}$, $h$) 
in units of the outermost galactic radius $R_g$, 
disk mass density ($\rho$) in units of
$M_g / R_g^3$ with $M_g$ denoting the total galactic mass, 
and rotational velocity [$V(r)$] in units of 
the characteristic galactic rotational velocity $V_0$.  
The disk thickness $h$ is assumed to be constant and small 
in comparison with the galactic radius $R_g$. 
Actually the physically meaningful quantity here is 
the combined variable $(\rho \, h)$ that represents 
the effective surface mass density on the thin disk.
As long as the disk thickness $h$ is much smaller than $R_g$,
its mathematical effect is inconsequential to the value of $(\rho \, h)$.
The gravitational forces of the finite series of concentric rings is described by the first term (double integral) while the centrifugal forces are described by the second term. 

In (\ref{eq:force-balance0}), we call the dimensionless paremeter 
$A$ ``galactic rotation parameter'', 
given by
\beq \label{eq:parameter-A}
A \equiv \frac{V_0^2 \, R_g}{M_g \, G} \, ,
\eeq
where $G$ denotes the gravitational constant, $R_g$ is the 
outermost galactic radius, and $V_0$ is the characteristic velocity 
(which is often chosen as the maximum asymptotic rotational velocity). 

According to (\ref{eq:force-balance0}), 
the corresponding rotational velocity $V(r)$ can be determined from  
a given distribution of $\rho(r)$ and the value of $A$, or vice versa.

Due the complexity of the equation, there are limited 
analytical formulas for a few simplified cases 
such as the Freeman disk and Mestel disk. 
Unfortunately neither of these simplified equations can 
satisfactorily describe the complete rotation curve 
given by (\ref{eq:smooth-V}).

By contrast, the numerical method presented by Feng \& Gallo \cite{FengGallo1}
solves an equation system including (\ref{eq:force-balance0}) and 
a conservation constraint for constant
total mass of the galaxy $M_g$, e.g.,   
\beq \label{eq:mass-conservation}
2 \pi \int_0^1 \rho(\hat{r}) h \hat{r} d\hat{r} = 1 \, ,  
\eeq
to determine $\rho(r)$ and $A$ from a given $V(r)$. 
Actually the same computational code can also be conveniently used to 
determine $V(r)$ and $A$ from a given $\rho(r)$, with a slight modification.
Thus, the methodology of Feng \& Gallo \cite{FengGallo1} 
represents a more general approach with the axisymmetric thin-disk model
to describe the galactic rotation for any measured rotation curve.

\section{Thin-Disk Models for Predicting Rotation Curves from 
Given Mass Distributions}
To facilitate a unified treatment for various thin-disk models,
we discretize the governing equations 
(\ref{eq:force-balance0}) and (\ref{eq:mass-conservation})
by dividing the one-dimensional problem domain $[0, 1]$
into a finite number of line segments called (linear) elements.
As described by Feng \& Gallo \cite{FengGallo1},
each element covers a subdomain confined by two end nodes,
e.g., element $i$ corresponds to the subdomain $[r_i, r_{i+1}]$
where $r_i$ and $r_{i+1}$ are the nodal values of $r$
at nodes $i$ and $i+1$, respectively.  
With each of the $N - 1$ elements mapped onto 
a unit line segment $[0, 1]$ 
in the $\xi$-domain (i.e., the computational domain),
$N$ independent residual equations can be obtained 
from the collocation procedure, i.e.,
\beq \label{eq:force-balance-residual}
\sum_{n = 1}^{N - 1} \int_0^1 \left[
\frac{E(m_i)}{\hat{r}(\xi) - r_i} - \frac{K(m_i)}{\hat{r}(\xi) + r_i}
\right] 
\rho(\xi) h \hat{r}(\xi) \frac{d\hat{r}}{d\xi} d\xi
+ \frac12 A V(r_i)^2
 = 0 \, ,
\eeq
where $K(m)$ and $E(m)$ denotes the complete elliptic integrals
of  the first kind and second kind, with
\beq \label{eq:mi-def}
m_i(\xi) \equiv \frac{4 \hat{r}(\xi) r_i}{[\hat{r}(\xi) + r_i]^2} \, .
\eeq

The $N$ residual equations (\ref{eq:force-balance-residual}) 
can be used to 
compute either the $N$ nodal values of $V(r_i)$ 
from given distribution of $\rho(r_i)$ or 
the distribution of $\rho(r_i)$ from a given set of $V(r_i)$,
with given values of $A$ and $h$.
Without loss of generality, the value of $h$ is assumed to be $0.01$
as comparable with that observed for the Milky Way galaxy.
If the constraint equation (\ref{eq:mass-conservation}) 
is also used with a discretized form
\beq \label{eq:mass-conservation-residual}
2 \pi \sum_{n = 1}^{N - 1} \int_0^1 
\rho(\xi) h \hat{r}(\xi) \frac{d\hat{r}}{d\xi} d\xi - 1 = 0 \, ,
\eeq
the value of $A$ can also be determined as part of 
the numerical solution.

These generally applicable equations are conveniently used 
for computing variables, even when analytical formulas are
available for some special cases.
Hence, a unified treatment for all cases is established 
for convenient comparison and analysis.
Moreover, as discussed by Feng \& Gallo \cite{FengGallo1},
imposing a boundary condition at the galactic center $r = 0$
for continuity of derivative of $\rho$,
i.e., in discretized form
\beq \label{eq:rho-1}
\rho(r_1) = \rho(r_2)
\, ,
\eeq
is desirable for obtaining high-quality numerical solutions consistent with reality.

\subsection{Freeman's Disk with Assumed Exponentially Decreasing Mass Distribution with Radius}
Freeman's (Ref.\cite{Freeman}) exponential disk gravitational model 
assumes a surface density function given by 
\beq \label{eq:exponential-density}
\rho(r) h = e^{-r / R_d}
\, , 
\eeq
where $R_d$ is a dimensionless scale length measured in units of $R_g$.
Freeman assumed the disk extends to infinity, 
so the solution can be written as 
\beq \label{eq:freeman-solution}
A \, V(r)^2 = 4 \pi R_d y^2 \left[I_0(y) K_0(y) - I_1(y) K_1(y)\right]  \, , 
\eeq
where $y \equiv r / (2 R_d)$
and $I_0$, $I_1$, $K_0$, $K_1$ are modified Bessel functions.

\begin{figure}[htb]
\resizebox{!}{1.25\textwidth}
{\includegraphics{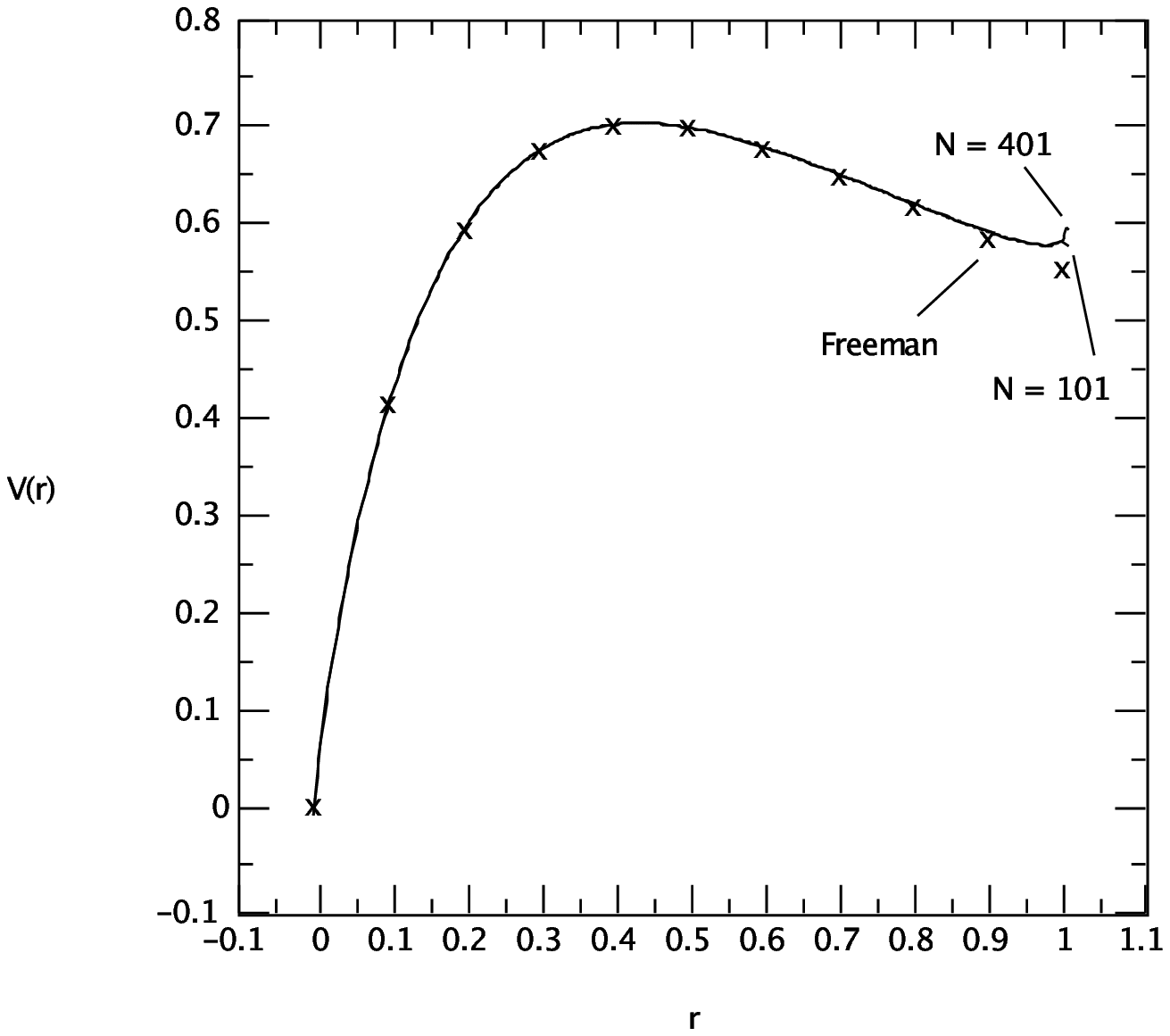}}
\caption{Rotational velocity $V(r)$ computed with
(\ref{eq:exponential-density}) for a given
surface density distribution $\rho(r) h = e^{-5 r}$
with $N = 101$ and $401$ are
represented by the solid curves.
The data points ($\times$) are calculated from Freeman's analytical solution
(\ref{eq:freeman-solution}) for an infinite disk with exponentially decreasing mass distribution. The computational codes are verified.}
\label{fig:Freeman}
\end{figure}

Figure \ref{fig:Freeman} shows the velocity $V(r)$ calculated 
with the present computational code, 
based on (\ref{eq:force-balance-residual}),
for an exponential disk truncated at $r = 1$ with $R_d$ set at $0.2$ 
and $A$ at $1.0$, and points marked as ``$\times$'' calculated with
(\ref{eq:freeman-solution}).
The comparison is generally quite good especially in the central core.
The slight discrepancy around the galactic rim $r = 1$ is expected 
due to the difference between truncated and infinite disks.
A slight difference between the cases with $N = 101$ and $401$
indicates that the
numerical discretization effects on the computed results
are insignificant here.

If the velocity profiles $V(r)$ shown in Figure \ref{fig:Freeman} 
for $N = 101$ and $401$ are used as known inputs,
the disk mass density $\rho(r)$ can be 
computed by solving
\[
\sum_{n = 1}^{N - 1} \int_0^1 
\rho(\xi) h \hat{r}(\xi) \frac{d\hat{r}}{d\xi} d\xi - 
\int_0^1 e^{- \hat{r} / R_d} \hat{r} d\hat{r} = 0 \, .
\]
For $R_d = 0.2$, 
\[
\int_0^1 e^{- 5 \hat{r}} \hat{r} d\hat{r} = 0.038383 \, .
\]
With a boundary condition,
\[
\rho(r_2) - \rho(0) = \frac{1 - e^{-5 r_2}}{5} r_2 \, ,
\]
imposed at $r = 0$, 
the computed disk mass density profiles,   
corresponding to the given
velocity profiles $V(r)$ with $N = 101$ and $401$,
both recover that given by (\ref{eq:exponential-density}) with $R_d = 0.2$. 
Moreover, the computed $M_g = 0.241167$ ($\approx 2 \times \pi \times 0.038383$),
and $A = 0.999768$ and $0.999987$, are respectively verified for $N  = 101$ and $401$.
Thus, the correctness of the computational code used 
herewith is verified.

The Freeman disk with an exponentially decreasing mass density
cannot produce a constant rotational velocity outside the central core, 
although the increasing velocity in the central core is reasonably described.

\subsection{Mestel's Disk with Assumed Mass Distribution Decreasing Inversely with Radius}
Another analytical model is Mestel's (Ref.\cite{Mestel}) disk,
which has an assumed mass density distribution that decreases inversely with radial position. 
The further assumption of finite disk yields the following equation with a termination function written as
\beq \label{eq:mestel-solution}
\rho(r) = \frac{A}{2 \pi h \, r} 
\left[1 - \frac{2}{\pi} \sin^{-1}(r)\right] \, ,
\eeq
to yield a constant velocity profile $V(r) = 1$.
However, the Mestel's disk has a mass density distribution
becoming singular at the 
galaxy center $r = 0$.  
This mathematical singularity at the galactic center is circumvented by
setting the nodal value of $\rho(r_1)$ equal to the value of $\rho(r_2)$,
equivalent to applying the boundary condition (\ref{eq:rho-1}).
With the rest of the nodes all given the $\rho$-values according to 
(\ref{eq:mestel-solution}) for $A = 1.572914$ (so that 
(\ref{eq:mass-conservation-residual}) is also satisfied), almost all
of the computed nodal values of $V(r)$ based on (\ref{eq:force-balance-residual})
are numerically equal to unity except for a few nodes around $r = 0$
with significantly reduced $V$-values
due to the boundary condition $(\ref{eq:rho-1})$.

This singularity problem is one undesireable feature that has limited the usefulness of Mestel's disk. And Mestel's incorrect description of the measured rotational profile in the galactic core is another reason for the limited applicability of this model. 
  
\subsection{A Model Based on Combination of Freeman and Mestel Disks} 

To reasonably capture the basic characteristics of 
an idealized galactic rotation curve (Eq.(\ref{eq:smooth-V})),
we combine the Freeman and Mestel disks, 
with Freeman describing the central core 
($r < R_c$) and Mestel outside the core ($r \ge R_c$).
In doing so, we assume that  
\beq \label{eq:FreemanMestel-solution}
\rho(r) = \left\{
\begin{split}
\rho_0 \, e^{- r / R_d} \, , \qquad \qquad \qquad 0 \le r < R_c
\\  
\frac{A}{2 \pi h \, r} 
\left[1 - \frac{2}{\pi} \sin^{-1}(r)\right] \, , R_c \le r \le 1 \,
\end{split}
\right . \, ,
\eeq
where
\[
R_d = 
\left\{\frac{1}{R_c} + \frac{2}{\pi \sqrt{1 - R_c^2} 
[1 - 2 \sin^{-1}(R_c) / \pi]}\right\}^{-1}
\]
and
\[ 
\rho_0 = \frac{A}{2 \pi h \, R_c e^{-R_c / R_d}} 
\left[1 - \frac{2}{\pi} \sin^{-1}(R_c)\right]
\, ,
\]
so that both $\rho$ and $d\rho / dr$ are continuous at $r = R_c$.
Moreover,
the mass conservation constraint (\ref{eq:mass-conservation-residual}) 
can also be satisfied by setting the galactic rotation parameter as
\[
A = \left[2 \, \pi \, \sum_{n=1}^{N-1} \int_0^1 \rho(\xi) h \hat{r}(\xi)
\frac{d \hat{r}}{d\xi} d\xi \right]^{-1} \, .
\]

\begin{figure}[htb]
\resizebox{!}{1.25\textwidth}
{\includegraphics{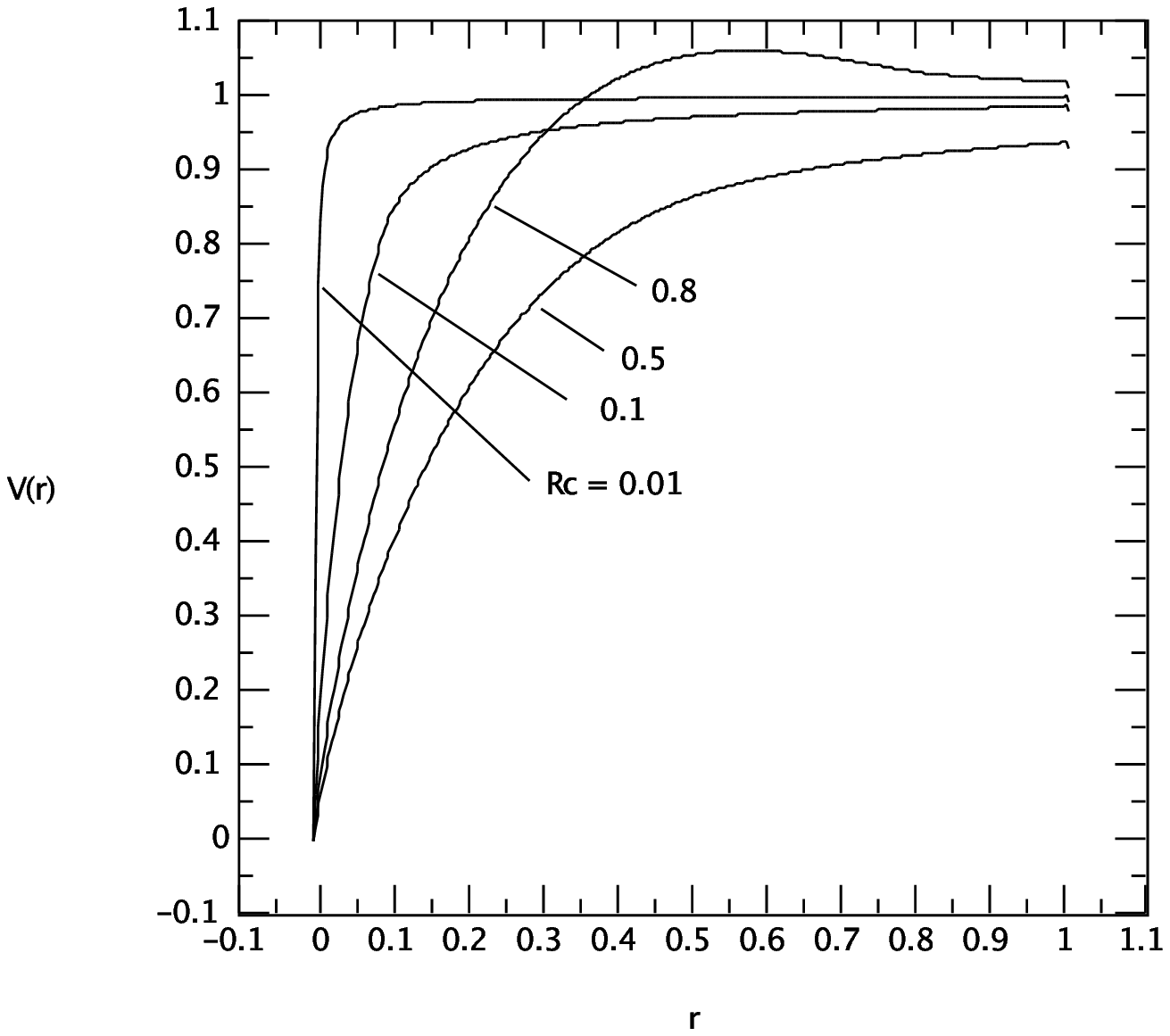}}
\caption{Velocity profiles $V(r)$ corresponding to
to the Combined Freeman/Mestel disks mass density distributions 
at different values of $R_c$ with $N = 401$.
The values of $A$ for $R_c = 0.01$, $0.1$, $0.5$,
and $0.8$ are computed as $1.5775$, $1.6424$,
$1.9304$, and $1.5096$, respectively.}
\label{fig:FreemanMestel-V}
\end{figure}

\begin{figure}[htb]
\resizebox{!}{1.25\textwidth}
{\includegraphics{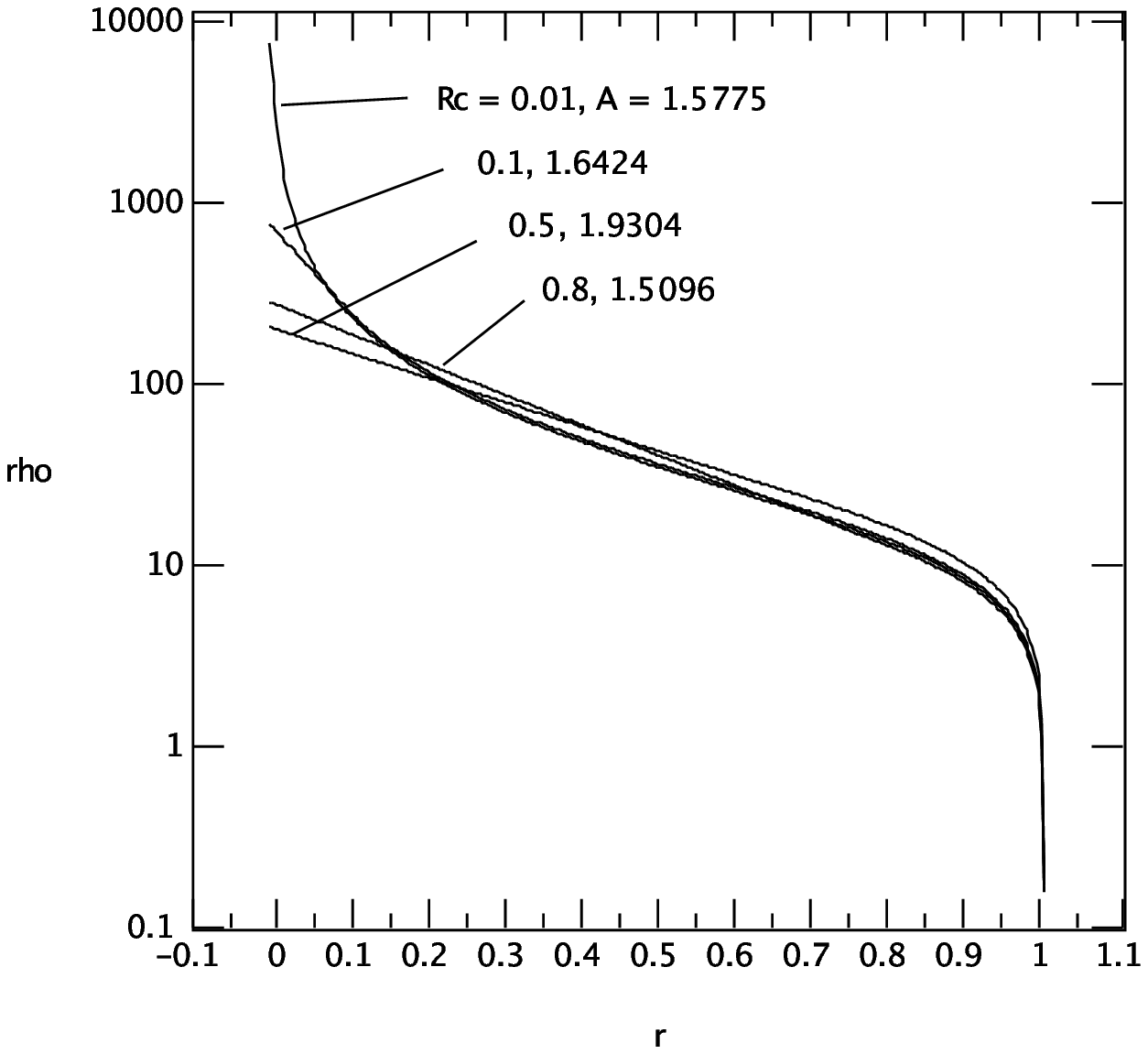}}
\caption{Selected density profiles $\rho(r)$ of
the Combined Freeman/Mestel disks corresponding to the velocity profiles
shown in Figure \ref{fig:FreemanMestel-V}
at different values of $R_c$ with $N = 401$.}
\label{fig:FreemanMestel-rho}
\end{figure}

Figure \ref{fig:FreemanMestel-V} shows
the rotation curves $V(r)$ for the Combined Freeman/Mestel disks at several values of $R_c$.
For $R_c < 0.6$ we find that $V(r)$ increases from $r = 0$ and 
approaches an asymptotic constant value in agreement with most galactic data.
But the velocity profile of Freeman disk alone dominates as $R_c$ approaches $1$
(e.g., $R_c = 0.8$).
The mass density profiles corresponding to the rotation curves in
Figure \ref{fig:FreemanMestel-V} are shown in 
Figure \ref{fig:FreemanMestel-rho}.
Freeman's exponential mass density curve dominates   
in the interval when $R_c \ge 0.5$,
except near the galactic rim $r = 1$ where the 
Mestel's formula takes over to force mass density to zero.

Pronko (Ref.\cite{pronko}) arrived at a model similar to the Freeman/Mestel combination from a different perspective. 

To summarize, the Freeman/Mestel combination disk yields reasonable descriptions of galactic rotation velocity profiles. However, with assumed mass distributions even the combination lacks the flexibiility and versatility of the Feng/Gallo model (Ref.\cite{FengGallo1}) (Eqs.2-4) which employs a computational search technique to determine the mass distribution that satisfies any measured rotational velocity profile accurately.

\section{Comparison with Thin-Disk Model for Numerically Predicting
Mass Distribution from Rotation Curve} 

\begin{figure}[htb]
\resizebox{!}{1.25\textwidth}
{\includegraphics{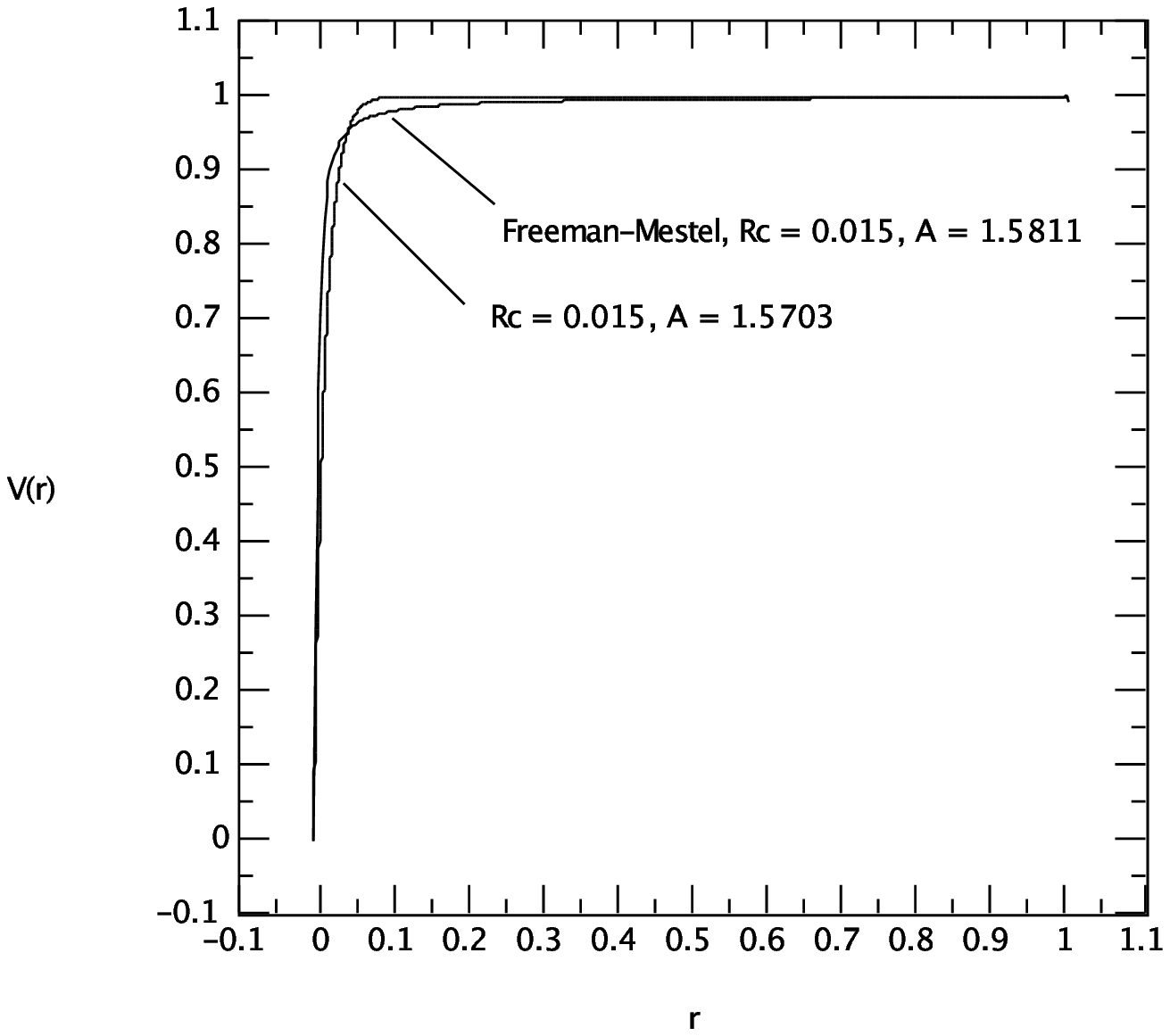}}
\caption{Reasonable comparison of $V(r)$ for $R_c = 0.015$ 
corresponding to $\rho(r)$ described by Combined Freeman/Mestel Disks  
(\ref{eq:FreemanMestel-solution}) 
and that computed from Feng \& Gallo \cite{FengGallo1} model (Eq.\ref{eq:smooth-V}).}
\label{fig:compare-V}
\end{figure}

\begin{figure}[htb]
\resizebox{!}{1.25\textwidth}
{\includegraphics{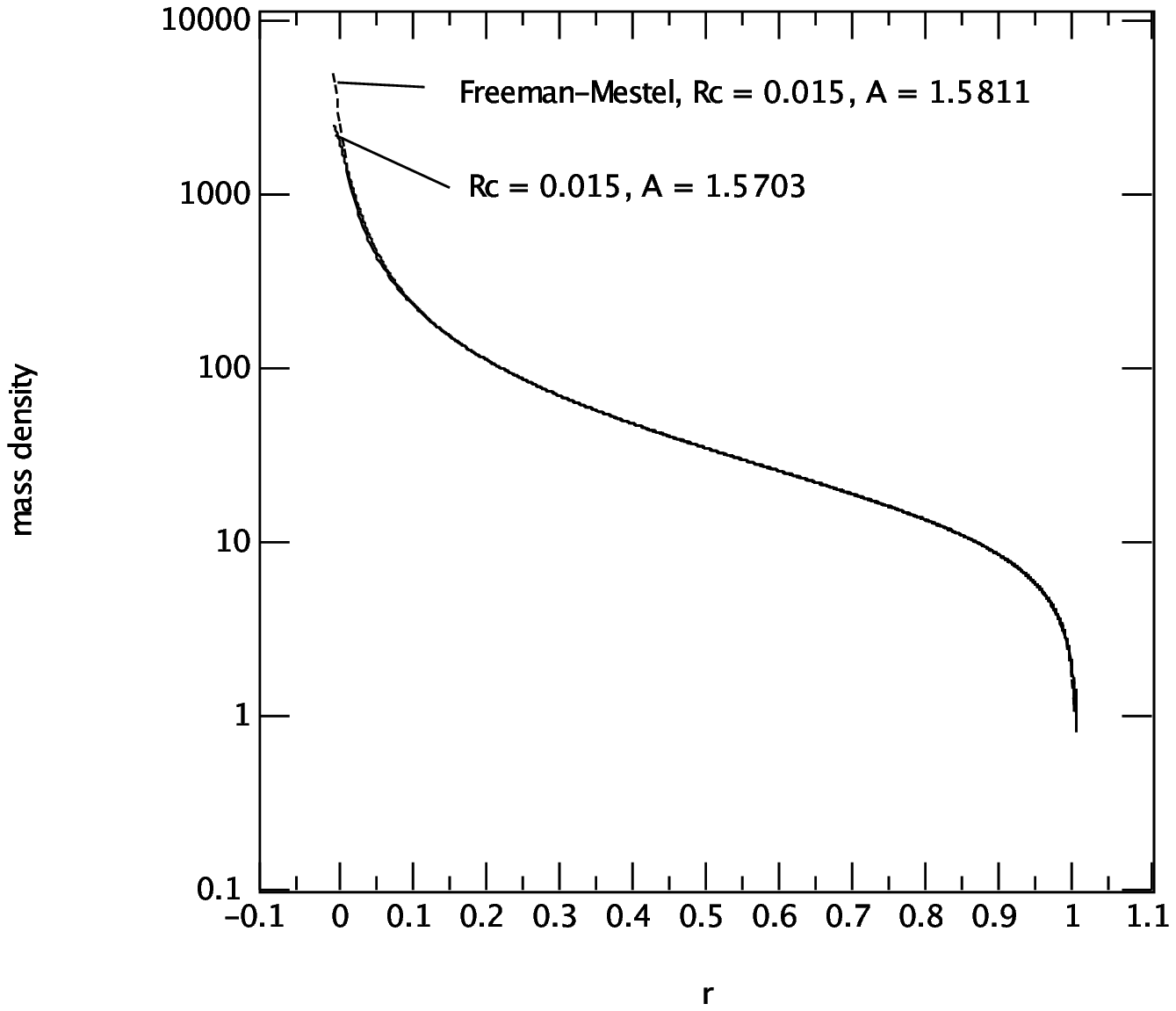}}
\caption{Reasonable comparison of $\rho(r)$ for $R_c = 0.015$ 
for Combined Freeman/Mestel disks  
(\ref{eq:FreemanMestel-solution}) (dotted curve) 
and that computationally predicted for the given rotation curve 
described by (\ref{eq:smooth-V}) (solid curve) and Feng \& Gallo \cite{FengGallo1} model.} 
The mass density decreases steeply from the galactic center, but then more slowly toward the periphery. 
\label{fig:compare-rho} 
\end{figure}

The rotation curves in 
Figure \ref{fig:FreemanMestel-V}
are not exactly the same as that described by (\ref{eq:smooth-V})
for a given value of $R_c$ simply because of 
the differences in mathematical definitions.
However, most of the curves are comparable.
For example, at $R_c = 0.015$ 
a mass distribution described by 
(\ref{eq:FreemanMestel-solution}) can produce a
rotation curve fairly close to that by (\ref{eq:smooth-V}) 
as shown in Figure \ref{fig:compare-V}.  
The corresponding mass density distributions 
are shown in Figure \ref{fig:compare-rho}.
As also found by Feng \& Gallo \cite{FengGallo1},
the mass density tends to decrease steeply 
from the galactic center, but beyond $R_c$, the mass density distributions decrease 
rather slowly towards the galactic periphery.  
This is consistent with the observed exponentially diminishing
luminosity data,
since both the temperature and opacity/emissivity 
are lower towards the periphery and we would expect the light distribution 
to decrease more quickly than that of mass towards the periphery.
We believe that
the assumption (often invoked by others) that the mass distribution 
exactly follows the light intensity distribution 
is fundamentally inaccurate due to lack of sound physical basis.
Our computational method for solving $\rho(r)$ based on observed $V(r)$
avoids that inaccurate assumption 
and yields rationally reasonable mass distributions 
corresponding to the measured galactic rotational profiles. 

\begin{figure}[htb]
\resizebox{!}{1.25\textwidth}
{\includegraphics{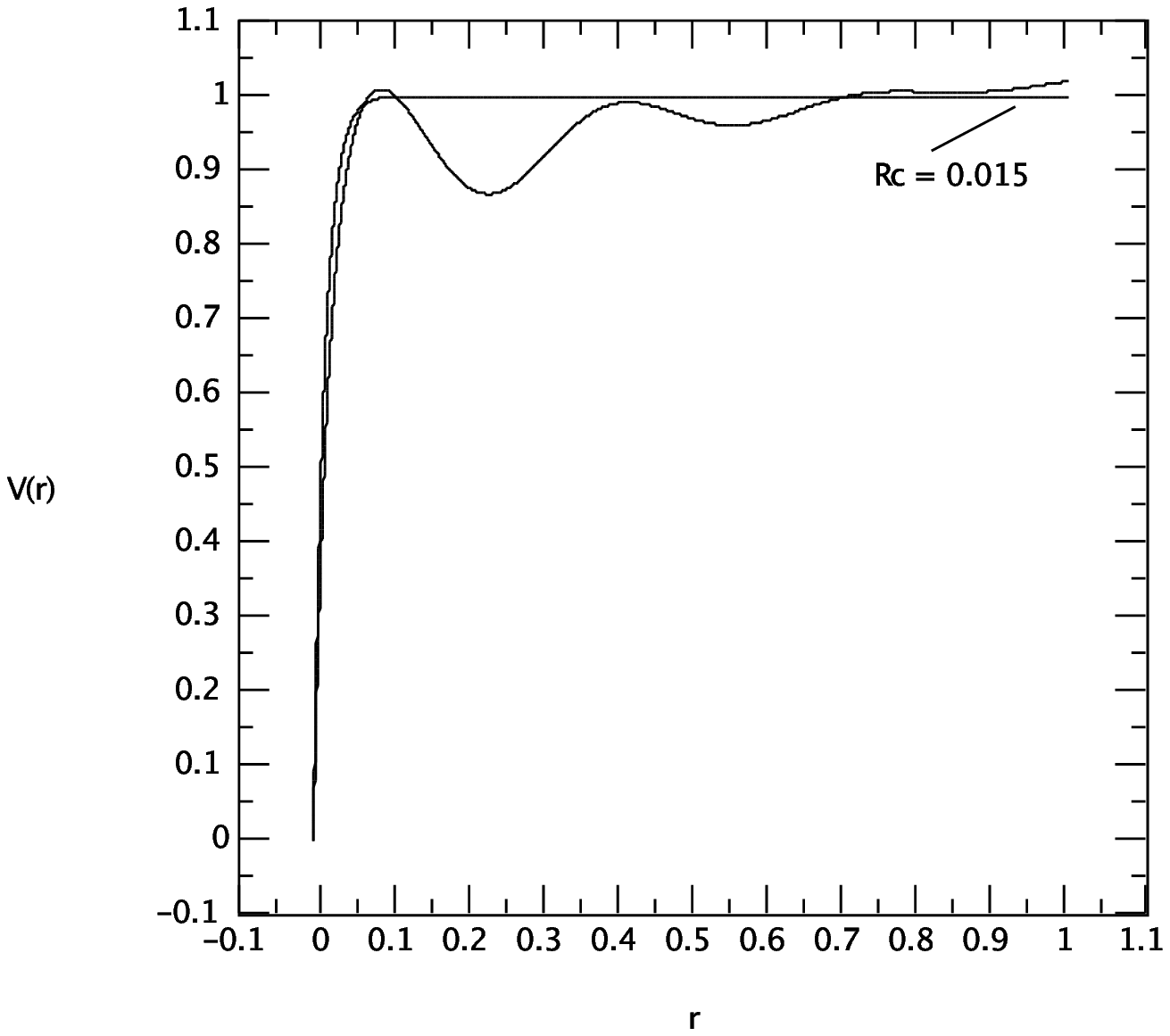}}
\caption{The more complex measured hypothetical wiggling rotation curve $V(r)$
described by (\ref{eq:wiggling-V}) together 
with that at $R_c = 0.015$ given by (\ref{eq:smooth-V}) (as labeled)
as a smoother idealized comparison. Actually, these curves are similar to the measurements of our own Milky Way Galaxy, and we use this information to calculate the total galactic mass of the Milky Way Galaxy in the text.}
\label{fig:velocity_r_2}
\end{figure}

\begin{figure}[htb]
\resizebox{!}{1.25\textwidth}
{\includegraphics{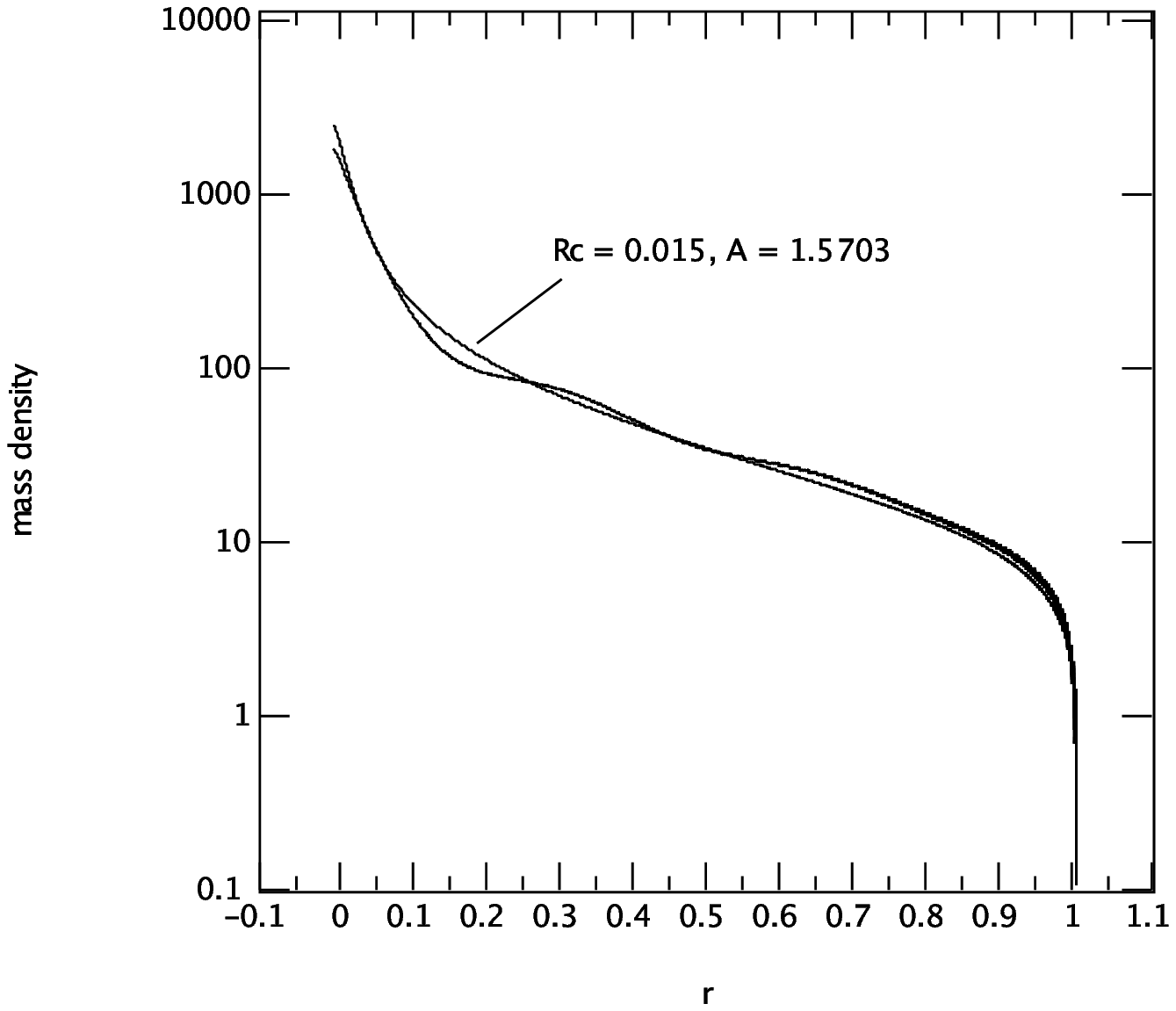}}
\caption{The more complex wiggling mass density distribution 
from the Feng \& Gallo \cite{FengGallo1} model
corresponding to the hypothetical wiggling rotation curve $V(r)$ 
(\ref{eq:wiggling-V}) compared with 
the smoother rotation curve given by (\ref{eq:smooth-V})
for $R_c = 0.015$.} The versatility of the Feng \& Gallo \cite{FengGallo1} model is apparent. 
\label{fig:density_r_2}
\end{figure}

Although the combined Freeman/Mestel disk model  
can produce reasonable rotation curves comparable to our 
idealized curves (Eq.\ref{eq:smooth-V})
especially for small $R_c$ (Figures \ref{fig:compare-V} and \ref{fig:compare-rho}), 
its capability is limited 
due to the presumed form of mass density distribution.  
In contrast, our Feng \& Gallo \cite{FengGallo1} model for numerically solving for 
$\rho(r)$ from a given $V(r)$ is much more 
versatile.
For example, measurements usually reveal a more complex 
wiggling rotation curve, 
as shown in Figure \ref{fig:velocity_r_2},
mathematically given by 
\beq \label{eq:wiggling-V}
V(r) = 0.85 \left[1 - e^{-r/0.02} + 0.2 \, e^{-4 \, r} \,
\sin(6 \, \pi \, r) + 0.2 \, \sqrt{r} \right] \, .
\eeq
The mass density distribution described by
(\ref{eq:FreemanMestel-solution}) cannot produce 
this more complex velocity profile. 
But from the computational Feng \& Gallo \cite{FengGallo1} model, 
the corresponding wiggling mass density distribution 
(Figure \ref{fig:density_r_2} 
and Eq.(\ref{eq:wiggling-V})) 
is computed by numerically solving 
(\ref{eq:force-balance-residual}) and (\ref{eq:mass-conservation-residual}).
The value of $A$ for the wiggling mass density curve is
computed as $1.5770$, as comparable to the reference value of $1.5703$.
When solving for $\rho(r)$ from a given $V(r)$, however, 
we need a much larger $N$ (e.g., $N = 4801$ used in 
the present work) than that for computing $V(r)$ from 
a given $\rho(r)$ (e.g., $N = 401$). This larger number of nodal points is needed 
for obtaining a good quality (smooth) curve with these more sharply varying input curves.

\section{Total Galactic Mass}
The measured rotational velocity profiles $V(r)$ 
includes knowledge of $V_0$ and $R_g$. 
With the computed value of the
galactic rotation parameter $A$, 
the total galactic mass
$M_g$ can be calculated as
\beq \label{eq:total-mass-Mg}
M_g = \frac{V_0^2 R_g}{A \, G} \, . 
\eeq
To check viability we investigate the idealized rotational 
velocity profile $V(r)$ of our own Milky Way galaxy shown in 
Figure \ref{fig:velocity_r_2} 
with $R_c = 0.015$. 
From the measurement data, the parameters appropriate for the Milky Way galaxy are $R_c = 0.015$, 
$V_0 = 2.5 \times 10^5 (m/s)$, and  
$R_g = 10^5 (\mbox{light-years}) = 9.46 \times 10^{20} (m)$. 
We obtain $A = 1.57$ for the mass density distributions shown in 
Figure \ref{fig:density_r_2}, 
and then the total galactic mass of the Milky Way is calculated  
from (\ref{eq:total-mass-Mg}) as
\[ 
M_g = 5.65 \times 10^{41} (kg) = 2.8 \times 10^{11} (\mbox{solar-mass})
\, .
\]   

This value is in very good agreement with Milky Way star counts of 100 billion, and further 
considering there is additional dust, grains, lumps, gases and plasma in all galaxies.  
In a subsequent companion publication, we shall employ  
Bulge+Disk models which are more compatible with visual observations of 
spiral galaxies and yield higher values of total galactic mass $M_g$. 
However, we emphasize the essential physics of galactic rotation is 
gravitationally controlled by the ordinary baryonic matter within 
thin galactic disks.  

Exploring solutions for a wide range of $R_c$ values,
Feng \& Gallo \cite{FengGallo1} noticed that
the computed values of $A$ are typically within a 
small range around $1.6$. 


\section{Ordinary Baryonic Matter versus Dark Matter}
The successful models (Freeman/Mestel Combo and Feng/Gallo) described in this paper explicitly assume that all galactic matter is contained within the galactic disk. And these models implicitly assume this is ordinary baryonic matter within the disk. Furthermore, the Freeman/Mestel Combo and Feng/Gallo models reproduce the measured galactic radial velocity profiles. Many other publications 
(Refs.\cite{pronko} - \cite{BalasinG})
employ similar computational techniques based on Newtonian gravity/dynamics (or General Relativity) similar to Feng/Gallo and also find radial mass distributions of ordinary baryonic matter within the galactic disk that satisfy the measured rotational velocity profiles.      
These mass distributions decrease roughly exponentially 
from the galactic center in the central core, 
but then decrease less rapidly (inversely with radius) 
towards the periphery. 
This decrease is slower than the measured light distribution. 
Thus there is ordinary baryonic matter within the galactic disk 
distributed towards the cooler periphery with 
lower emissivity/opacity and therefore darker. 
There are no mysteries in this rational scenario based on verified physics.   

However, there are many other publications (Refs.\cite{OstrikerPY}-\cite{deJong2}) that take a different track. 
Analogous to the measured radial light distribtuion, they start with a radial mass distribution within the disk that decreases approximately exponentially with radius. This is called the Mass/Light ratio assumption. But these models do not yield the measured rotational velocity profiles. 
To compensate, speculations are invoked re the existence of ``massive peripheral spherical halos of mysterious Dark Matter'' around the galaxies. 
But no significant matter has been detected in this 
untenable unstable gravitational halo distribution. 
This speculated Dark Matter is ``mysterious'' since it does not 
interact with electromagnetic fields (light) nor ordinary matter 
except through gravity. This Dark Matter must have other 
abnormal (non-baryonic) properties to maintain 
its peripheral spherical shape against the galactic rotation 
and gravitational attraction of ordinary matter. 
Furthermore, the assumption of a simple proportional relationship between mass and  
light intensity is not based on convincing physical principles. It does not accurately describe the 
the strong influence of both temperature and opacity/emissivity on light intensity. 
These deficiencies become apparent from edge-on views of 
spiral galaxies with a dark galactic line
against a bright galactic background, 
revealing the substantial radial temperature and opacity/emissivity gradients across the galaxy. 
The concept of mass/light ratio is too inaccurate and oversimplified to apply to a complex galactic structure. Many unverified ``mysteries'' are invoked as solutions 
to real physical phenomena. 

To explain the failures associated with the assumption of the mass/light ratio, a different speculation has been proposed (Ref.\cite{Milogram}) re deviations from Newtonian gravtiy/dynamics. 
But there is no independent experimental evidence of such deviations. 
Our use of Newtonian gravity/dynamics with sound computational techniques 
has proven sufficient to explain the observed rotational velocity curves.   

In conclusion, our approach utilizing Newtonian gravity/dynamics 
to computationally solve for the ordinary baryonic mass distributions 
within the galactic disk simulates reality and agrees with 
observed data.

\section{Limitations and Strengths of Thin Disk Model}
Our simple thin disk model does not address many important features 
such as spiral structure, plasma effects, galactic formation, 
galactic evolution, galactic jets, black holes, relativistic effects, 
galactic clusters, etc.. 

It is well known (Ref.\cite{BT}) that the internal gravitational behavior 
of a thin disk is much different than a sphere.  
This distinctly different behavior enables our thin disk model 
to describe the rotational dynamics of mature spiral galaxies, 
and the total galactic mass.  

Our thin disk model has finite radial extent. 
Beyond the galactic radius, we assume the density has dropped to 
the inter-galactic level, which is approximately spherically symmetric 
and thus no longer affects the galactic dynamics. 
We mention this because some others (Ref.\cite{BT}) have taken 
the relevant integrals to infinity, which we think is inappropriate. 

In our thin disk approach, we balance the gravitational forces against 
the centrifugal forces at each and every point. 
Thus, our solutions for the mass distributions and total galactic mass 
satisfy the measurements and ensure stability within the same context 
as similar calculations for our Solar System and Earth satellites. 
Some previous authors obtain solutions that are not gravitationally 
stable because they obtain incorrect mass distributions and incorrect 
galactic masses and do not satisfy the measured rotational profiles. 
Thus, their solutions are unstable, whereas our solutions are stable 
within the Newtonian context.  

Plasma effects are certainly active in the formation and 
evolution of galaxies from the original hot 
plasma (Refs.\cite{Peratt1}-\cite{Peratt3}). However, for the mature 
spiral galaxies we are addressing, the free plasma density has 
dropped to levels sufficiently low that plasma does not affect 
the predominantly gravitational galactic dynamics. 
This is evidenced in our own Solar System in which gravitational dynamics 
dominate even in the presence of solar wind, coronal mass ejections, 
comet tails, etc. The plasma in our Sun is stabilized by gravitational forces, 
even though plasma effects are very active within the Sun itself. 
Since our Solar System is approximately 1/3 distance from our 
Milky Way galactic center, we have our Solar System evidence for 
the dominance of gravitational forces within our own Milky Way galaxy, 
at least out to these radial distances.      

Again, we repeat, our thin disk model is sufficient 
to describe the rotational dynamics of mature spiral galaxies 
and the total galactic mass. 

\section{Conclusions} 
The measured rotation velocity profiles of mature spiral galaxies are successfully described with an appropriate combination of Freeman and Mestel thin disk gravitational models. 
These models start with assumed analytical radial mass distributions. 

By contrast, our approach utilizes Newtonian gravity/mechanics to computationally solve for radial mass distributions that satisfy the measured rotational velocity profiles. 

All these results reproduce the measured rotational velocity profiles, even though the fundamental approaches are quite different. 

Also, all the radial mass distributions within the galactic disk are also mutually compatible. 
We compute that in the central galactic core, the mass density decreases approximately exponentially (similar to Freeman model). 
But beyond the core out to the periphery, we compute that the  mass density decreases more slowly (inversely with radius similar to Mestel model).   
It is important to realize this mass distribution toward the galactic periphery is darker due to the lower temperature and lower opactiy/emissivity in these outer regions. This is apparent from edge-on views of galaxies which display a dark disk-line against a much brighter galactic halo.  

Most other previous research assumes a galactic density decreasing exponentially with radius out to the galactic periphery, analogous to the  measured light distribution. But this assumption (by others) is inaccurate since both the temperature and opacity/emissity are important but ignored variables. There is no simple relationship between mass and light. These prior models do NOT describe the measured velocity profiles, and speculations are invoked re halos of mysterious Dark Matter or gravitational deviations to compensate. The Dark Matter must have ``mysterious'' (non-baryonic) properties because there is no evidence of its existence and it is not responding to gravitational, centrifugal and electromagnetic forces in any known manner. 
By contrast, our results indicate no massive peripheral spherical halos of mysterious Dark Matter and no deviations from simple gravity. Our total galactic mass determinations are also in agreement with data.   

The controversy is summarized as follows. 

We believe there is ordinary baryonic matter within the galactic disc distributed more towards the galactic periphery which is cooler with lower opacity/emissivity (and therefore darker). 

Others believe there are massive peripheral spherical halos of mysterious Dark Matter surrounding the galaxies.

\section{Acknowledgements}
We gratefully acknowledge Louis Marmet, Ken Nicholson and Michel Mizony whose intuition and computational techniques convinced us that galactic rotation could be described by suitable mass distributions of ordinary baryonic matter within galactic disks. 
Anthony Peratt originally sparked our interest with his plasma dynamical calculations re the formation and evolution of galaxies. Ari Brynjolfsson has energetically supported our efforts.


\end{document}